\documentclass{eas}
\usepackage{graphicx,txfonts,natbib}
\usepackage[utf8]{inputenc}

\begin{document}

\title{High Resolution Laboratory Spectroscopy}
\author{S. Brünken}
\address{I. Physikalisches Institut, Universität zu Köln, Zülpicher Str. 77, 50937 Köln, Germany}
\author{S. Schlemmer}
\sameaddress{1}
%
%

\begin{abstract}
In this short review we will highlight some of the recent advancements in the field of high-resolution laboratory spectroscopy that meet the needs dictated by the advent of highly sensitive and broadband telescopes like ALMA and SOFIA. Among these is the development of broadband techniques for the study of complex organic molecules, like fast scanning conventional absorption spectroscopy based on multiplier chains, chirped pulse instrumentation, or the use of synchrotron facilities. Of similar importance is the extension of the accessible frequency range to THz frequencies, where many light hydrides have their ground state rotational transitions. Another key experimental challenge is the production of sufficiently high number densities of refractory and transient species in the laboratory, where discharges have proven to be efficient sources that can also be coupled to molecular jets. For ionic molecular species sensitive action spectroscopic schemes have recently been developed to overcome some of the limitations of conventional absorption spectroscopy. Throughout this review examples demonstrating the strong interplay between laboratory and observational studies will be given.
\end{abstract}
\maketitle
\section{Introduction}

Our understanding of the physics and chemistry of the interstellar medium (ISM), and in particular of star formation processes, has increased considerably in recent years, as the many contributions to these proceedings impressively demonstrate. Molecular rotational absorption and emission lines in the mm-wavelength to FIR regime are the main tracers of physical and chemical conditions in a wide variety of astronomical environments, ranging from the earliest stages of star formation such as dense prestellar clouds, via extremely line-rich high-mass star-forming regions, all the way to circumstellar shells around late-type stars that are feeding back molecular material to the interstellar medium.\\

\begin{figure}
   \centering
   \includegraphics[width=\hsize]{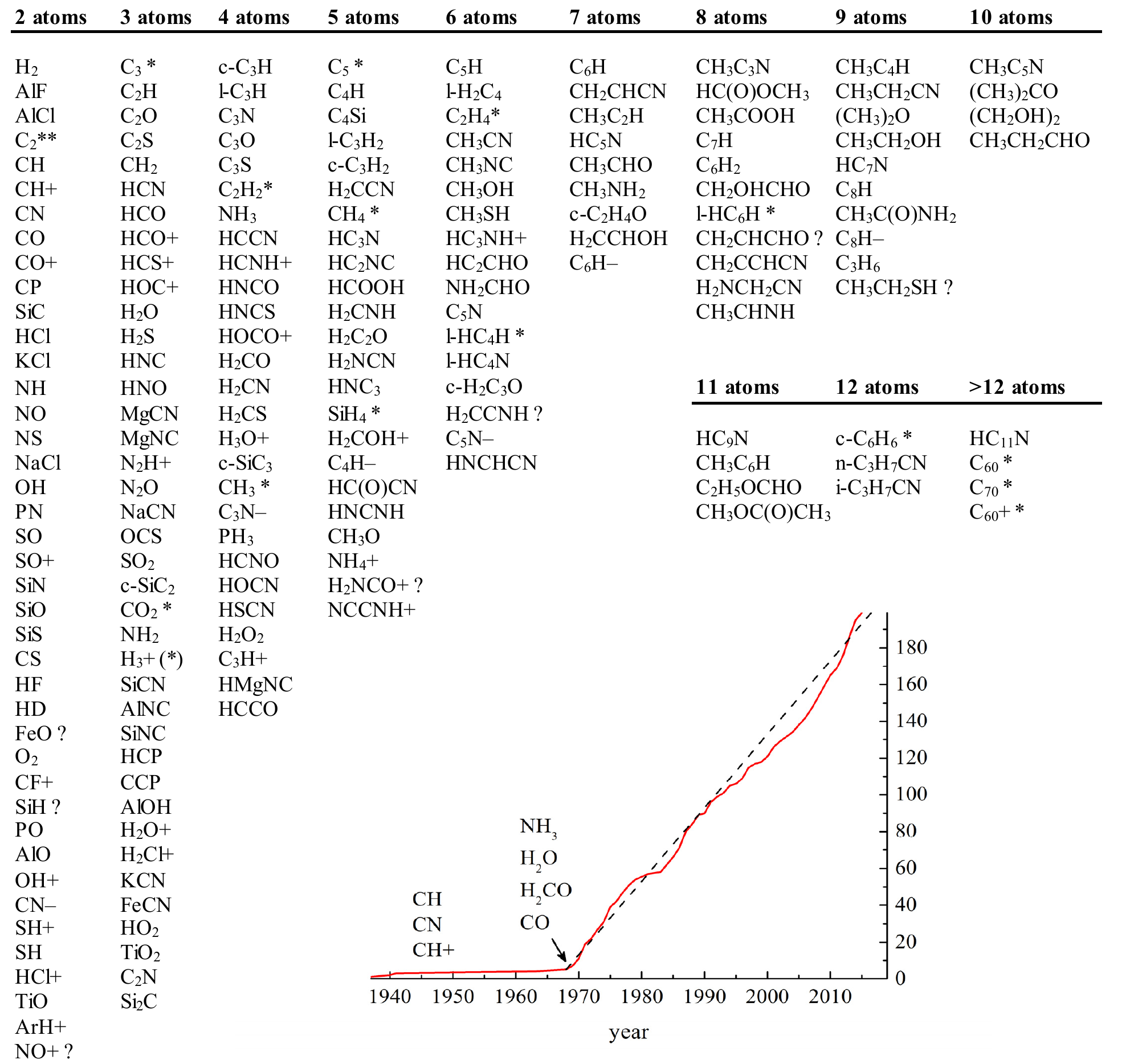}
      \caption{The nearly 200 molecules detected in the inter- and circumstellar medium (listed in the CDMS (Cologne Database for Molecular Spectroscopy) as of Sept. 15, 2015; $*$, $**$  denotes species detected via their vibrational or electronic spectrum, resp.). The insert shows the steady increase of identified molecules at a rate of around 4 new detections per year \cite[updated figure from][]{Tha2006}.}
         \label{ismmols}
\end{figure}

 To date, nearly 200 molecules have been detected in the inter- and circumstellar medium (see Fig.\ref{ismmols}). With the advent of radio-astronomy in the late 1960's, most of these species were identified by their rotational transition lines, with an average of around four new detections per year (see inset in Fig.\ref{ismmols}). This trend is not expected to abate owing to ever more sensitive observatories, like ALMA and NOEMA operating currently, and the expansion of the accessible frequency range, e.g. with SOFIA and formerly Herschel towards the THz/FIR regime. The extension of the list of known molecular species should by no means be viewed as some sort of interstellar ``stamp collection''. Instead, many of the recent detections have proven to be valuable molecular probes to study certain aspects of the physics and chemistry in a variety of astronomical environments. One example (among many others) is the observation of the argonium ion ArH$^+$ with the Herschel satellite, first detected in a supernova remnant \citep{BSO2013}, and successively in the diffuse galactic and extragalactic interstellar medium \citep{SNM2014, MMS2015}, where it can serve as a tracer for almost purely atomic gas. Similarly, interferometric observations with the SMA and ALMA of rotational transitions of the refractory titanium dioxide molecule (TiO$_2$) can tell us about the first steps of dust formation in oxygen-rich circumstellar envelopes \citep{KGM2013, DVM2015}. Furthermore, an understanding of the astrochemistry of molecular clouds requires an as complete as possible chemical inventory, in particular towards increasing molecular complexity \citep[see, e.g., ][for reviews]{EC2000,HD2009,Tielens2013}. One of the main question concerns the formation pathways of  large organic molecules that are observed in the interstellar medium \citep[e.g.,][]{LZS2013,BGM2014}, and their possible role in the formation of prebiotic molecules.\\

In all these areas, the detection and identification of new molecular species and the interpretation of the astronomical observations rely on accurate spectroscopic data on transition frequencies and intensities provided by high-resolution laboratory studies. The laboratory spectroscopist's wishlist mirrors in many respects that of the observational  astrophysicist, and often profits in a similar way from technological developments in receiver technology. It includes: high sensitivity, large instantaneous bandwidth coupled with high spectral resolution, and the exploration of new frequency bands (e.g., THz/FIR). An additional challenge in the laboratory is to establish efficient synthesis pathways to produce sufficient amounts of a ``new'' molecular species in the gas phase.\\

In this short review we want to highlight some of the recent advances in the field of laboratory spectroscopy covering exemplary the different aspects mentioned above.

\section{Broadband techniques to explore molecular complexity}

When we talk about complex organic molecules (COMs) in the context of astrochemistry, they are still rather small from the perspective of an organic chemist, consisting of typically $6-15$ atoms. Nevertheless, their rotational spectra are in fact quite complex, owing to their generally small rotational constants and the presence of permanent electric dipole moment components along all three principal axes of inertia. Further complications arise in the case that the molecule comprises  internal rotors (e.g. a methyl side group), and due to energetically low-lying vibrational states. In order to determine an accurate spectroscopic model, i.e. spectroscopic constants from which transition frequencies and intensities can be predicted to compare to astronomical observations, it is, therefore, often necessary to measure several thousands of lines in the laboratory over a broad range from the microwave up to the THz frequency region.

\subsection{\label{abs} Absorption spectroscopy}

Conventional absorption spectroscopy still is the workhorse in high-resolution rotational laboratory spectrocopy. An absorption spectrometer consists of an intense, tunable, monochromatic radiation source, an absorption cell containing the gas sample, and a sensitive broadband detector (e.g., room temperature Schottky diodes or liquid helium cooled hot electron bolometer). Solid state multiplier configurations have in recent years prevailed as standard radiation sources over, e.g., backward wave oscillator tubes \citep[described in, e.g., ][]{WKT1994,PGB1997}. They rely on the up-conversion of a fundamental frequency provided by, e.g., a commercial synthesizer, by a cascade of non-linear multiplier devices (most commonly planar Schottky diodes). The technology is driven mainly by local oscillator demands for heterodyne receiver development, and multipliers and complete multiplier chain configurations covering the $50 - 2000$~GHz range \citep[and some windows above,][]{CHR2011} are nowadays commercially available. Due to their high accuracy ($\Delta \nu < 1$~kHz), large scanning bandwidth, high output power (several 10 $\mu$W to mW), and easy handling they are perfectly suited for broadband high-resolution spectrocopic studies \citep{DMP2005,DKM2014}.\\

\begin{table}
\caption{\label{table_coms}Exemplary recent laboratory studies of complex organic molecules in the mm- and submm-wavelength range using absorption spectrometers based on solid state multiplier sources.}
{\footnotesize
\centering
\begin{tabular}{lllll}
\hline\hline
Species 			&  	frequency 	& reference 		& detected 		 	& laboratory\\
	  			&	range [GHz] 	&			& in ISM     				&	\\
\hline
d-CH$_3$CH$_2$OH	& $35 - 500$		&\cite{WSO2015} 	&main isotopologue		&Cologne\\
$^{13}$C-CH$_3$CHO	 & $50 - 945$	& \cite{MMI2015} 	& main isotopologue	& Lille \\
CH$_3$NH$_2$		& $500 - 2600$ 	&\cite{MID2014}	&\cite{GBR1973}		& Lille, JPL \\
CH$_2$DOCH$_3$ 		& $150 - 990$ 	& \cite{RMC2013} 	& \cite{RMC2013} 	&Lille\\
$^{13}$CH$_2$OHCHO     & $150 - 945$	& \cite{HMM2013} 	&no				&Lille\\
$^{13}$C-(CH$_3$)$_2$O & $70 - 1500$ 	&\cite{KBE2013}	&\cite{KBE2013}		&Cologne\\
HCOOCH$_2$D  		&  $770 - 1200$	& \cite{CDT2013}	&\cite{CDT2013}		&JPL\\
DCOOCH$_3$ 		& $850 - 1500$	& \cite{DCY2015}	&main isotopologue		&JPL\\
$^{13}$C-CH$_3$CH$_2$OH & $80 - 800$ 	&\cite{BWM2012} 	& main isotopologue 	&Cologne \\
CH$_3$CHO, $\nu_t=0,1,2$ & $50 - 1620$	& \cite{SAI2014}	& ground state		& Lille, JPL\\
HCOO$^{13}$CH$_3$ $\nu _t=0,1$ & $50-940$ & \cite{HCT2014} & \cite{HCT2014}	 	& Lille\\
CH$_2$CHCOOH 		&$95 - 397$ 	&\cite{CMD2015} 	& no				& Bologna\\
CH$_2$CHCHO		& $50 - 660$ 	& \cite{DBK2015} 	& \cite{HJL2004} 		& Valladolid\\
CH$_2$CHCOOH		& $130-360$ 	& \cite{AKP2015} 	& no 				& Valladolid \\
CH$_3$OCHCH$_2$ 	& $50 - 650$ 	& \cite{DKM2014} 	&no 				& Valladolid\\
CH$_3$CHOHCH$_2$OH	& $38-400$		&\cite{BOM2014} 	& no				&Cologne \\
CH$_3$COCH$_2$CH$_3$ &  $73 - 1000$ 	& \cite{KSW2014} 	& no 		 		& Emory \\
H$_2$CCHCH$_2$NC 	& $150 - 900$	& \cite{HMH2013}	&no				&Lille\\
CH$_2$NCH$_2$CN 	& $120 - 600$ 	& \cite{MMG2013} 	&no				&Lille\\
C$_5$H$_6$O 		& $75 - 970$ 	& \cite{FSW2012} 	& no 		 		& Emory \\
n-C$_4$H$_9$CN		& $75 - 348$ 	&\cite{OMW2012} 	& no				&Cologne	\\

\hline
\end{tabular}}
\end{table}

In the past years several laboratories worldwide have provided a wealth of rotational spectroscopic data on complex organic molecules relevant for astrochemistry using these kind of absorption spectrometers at mm- to submm-wavelengths. As the examples listed in Table \ref{table_coms} demonstrate, these studies targeted on the one hand isotopologues and vibrationally excited states of already detected species. Some of these have already been observed in star-forming regions (the ``molecular factories'') in the ISM, and many more are likely observable and need to be identified in high sensitivity line surveys with ALMA. In many cases the isotopic species may be used as tracers on possible formation routes as, e.g., in the case of $^{13}$C-substituted dimethyl ether \citep{KBE2013}. \\

\begin{figure}
   \centering
   \includegraphics[width=\hsize]{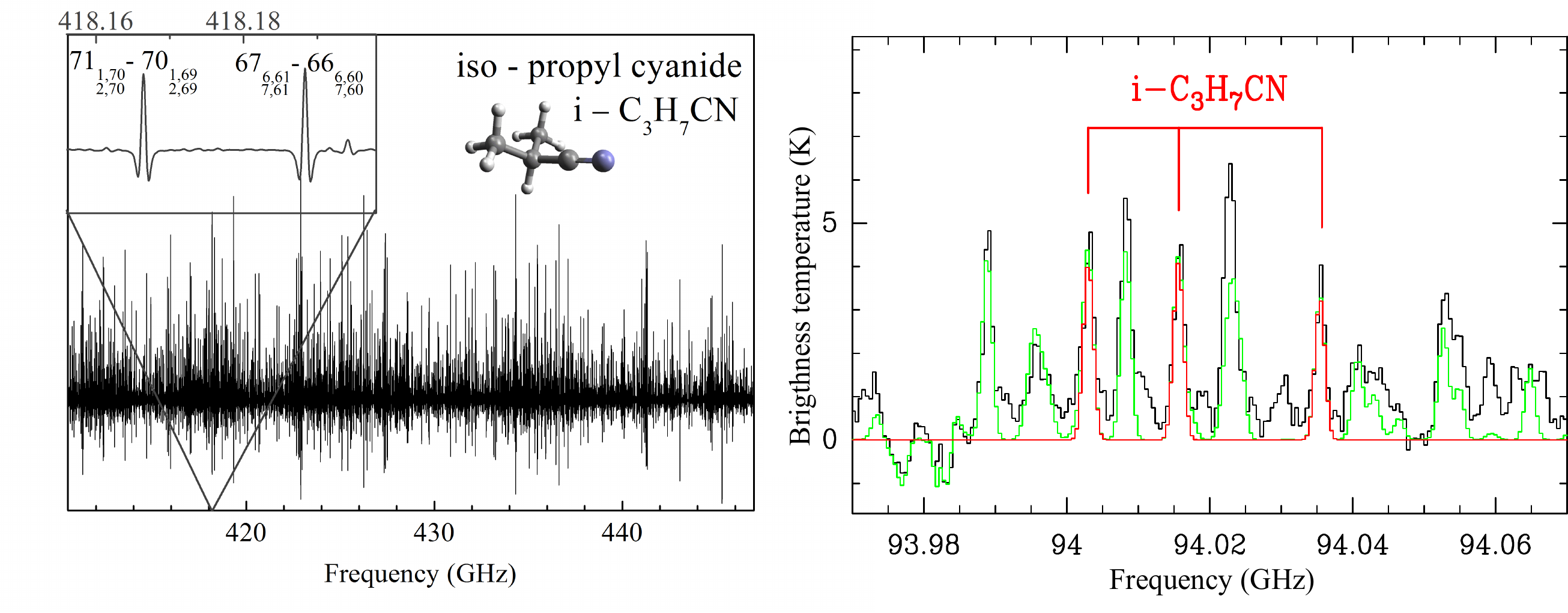}
      \caption{First detection of a branched carbon-backbone molecule in the interstellar medium: overview spectrum of the complex organic molecule iso-propyl cyanide recorded with a solid-state multiplier device based absorption spectrometer in Cologne (left), covering a frequency bandwidth of $>30$~GHz (H. M{ü}ller, private communication, extension of spectroscopic data in \cite{MCW2011}). High frequency resolution can be achieved, as seen in the inlay, covering a much narrower frequency range. Several transitions of the same molecule observed with ALMA towards Sgr B2(N) are displayed on the right \citep{BGM2014}.}
         \label{iPrCN}
\end{figure}

Other studies targeted species that are chemically linked to known astronomical COMs, providing accurate transition frequencies for their identification in the ISM. A very nice example demonstrating the strong interplay between laboratory spectroscopists and observational astronomers is the recent detection of {\it iso}-propyl-cyanide, (CH$_3$)$_2$CHCN, towards the star-forming region Sgr B2(N) within the EMoCA (Exploring Molecular Complexity with ALMA) project \citep{BGM2014}, based on preceding high-accuracy broadband laboratory studies \citep{MCW2011}, see Fig \ref{iPrCN}. This detection marks the first observation of an organic molecule with a branched carbon backbone, a structure also found in amino acids. Spectroscopic data on the next larger member of the alkyl cyanide series, {\it n}-C$_4$H$_9$CN, also existing in branched variants, is already available \citep{OMW2012}.

\subsection{Chirped Pulse Fourier Transform Microwave Spectroscopy}

\begin{figure}
   \begin{minipage}[c]{0.57\textwidth}
   \includegraphics[width=\textwidth]{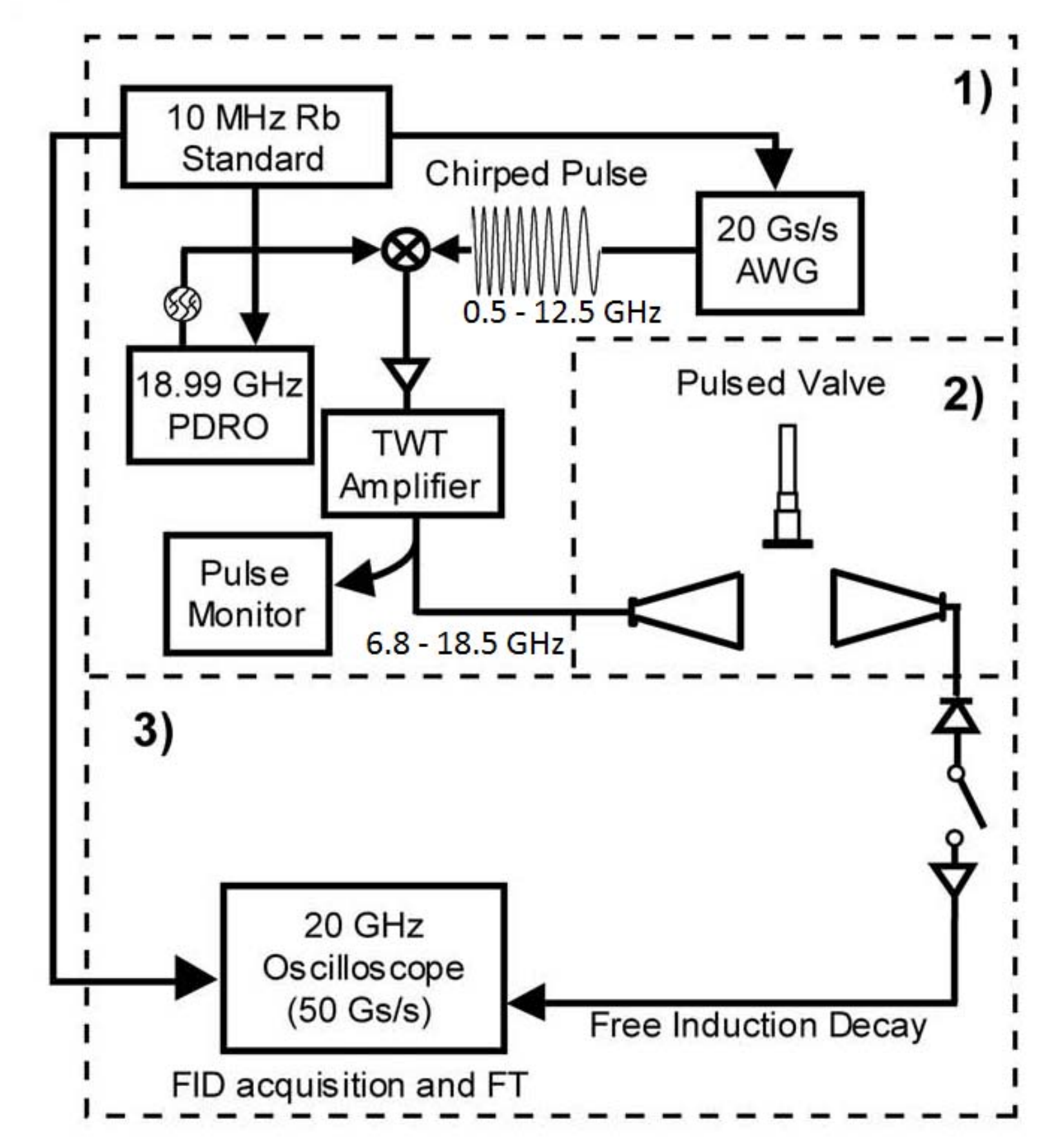}
   \end{minipage}\hfill
   \begin{minipage}[c]{0.4\textwidth}
      \caption{Schematic of a chirped pulse Fourier transform microwave spectrometer introduced by \cite{BDD2008}. A 12 GHz bandwidth (1~$\mu$s long) microwave pulse is generated by an arbitrary waveform generator (AWG), upconverted to the range $6.8-18.5$~GHz and amplified to several 100~W (1), before entering the molecular beam chamber (2). The coherent free induction decay (FID) of the excicted molecules, containing the spectral information, is then recorded and analysed with a fast digital oscilloscope. (Reprinted with permission from \cite{BDD2008}. Copyright 2008, AIP Publishing LLC.)}
         \label{chirp}
         \end{minipage}
\end{figure}

Since its development by Brooks Pate's group in Virginia \citep{BDD2006,BDD2008} chirped-pulse Fourier Transform Microwave (CP-FTMW) spectroscopy has proven to be a powerful alternative to conventional absorption spectroscopy (see Fig. \ref{chirp} for experimental details). CP-FTMW is a truly broadband method, allowing the acquisition of a complete spectrum (with typically 10~GHz bandwidth) in a matter of several 10 $\mu$s, making it also very sensitive compared to scanning absorption techniques by integrating this multiplexed signal over several single acquisitions.
Originally developed for the radio-band ($<20$~GHz), the method has been extended by several groups for use up to the 500~GHz region \citep{PSK2011,GDP2011, ZNM2012,SHN2012}. Its applications include trace gas sensing \citep{GDP2011}, structure determinations of biomolecules \citep{MPC2012,BZS2015} and of weakly bound clusters \citep{SKW2014,TSJ2014}, investigation of reaction dynamics \citep{AZP2014,ZSW2014}, and of molecular chirality \citep{PD2013}, and of course it is widely used for high resolution rotational spectroscopy of complex organic and other molecules of astrophysical interest \citep[e.g.,][]{CMM2013,BMZ2013,KSW2014,AKP2015}. \\

One particularly interesting example making use of the broadband capability of CP-FTMW comes from a collaborative study of the Harvard and Virginia laboratory astrophysics group with the Green Bank Telescope (GBT) PRIMOS survey project \citep{HRJ2011}. In what the authors call a "gas-phase Miller-Urey experiment", they recorded broadband ($9 - 50$~GHz) laboratory spectra of CH$_3$CN and NH$_3$ (or H$_2$S) discharge reaction products seeded in a molecular beam. After weeding out lines from known species, they identified and spectroscopically analysed new complex molecules with the help of quantum chemical calculations. By comparing this new data with the GBT survey data towards Sgr B2(N), two new species could unambiguously be detected: E-cyanomethanimine (HNCHCN) and ethanimine (CH$_3$CHNH), possible precursor molecules to the nucleobase adenine and the amino acid alanine, respectively \citep{LZS2013,ZSS2013}.

\section{Spectroscopy at Terahertz frequencies}

The extension of the accesible frequency domain for high-resolution laboratory spectroscopy into the Terahertz (or FIR) range is another vivid field of research. Many light hydrides, containing a heavy element atom and one or more hydrogen atoms, have their lowest-lying rotational transitions at frequencies near or above 1~THz. As \cite{GNG2016} point out in a recent review, their observation with Terahertz observatories like Herschel (HIFI, PACS, SPIRE) and SOFIA (GREAT, FIFI-LS) have and will be used as diagnostic probe in many interstellar environments \citep[e.g.,][reporting on observations of CH, NH, H$_3$O$^+$, SH$^+$, H$_2$O$^+$, SH, HCl$^+$, OD, and  p-H$_2$D$^+$]{BBD2010,NFG2012,DGN2012,PDL2012,BSC2014,AUL2014}. Furthermore, low-lying vibrational modes of complex molecules, like carbon-chains \citep{MGS2010} and polycyclic aromatic hydrocarbons (PAHs) fall into the FIR region, possibly better suited for the identification of a specific PAH in space than their ubiquituous mid-infrared emission features.

Frequency up-conversion with commercial solid-state multiplier devices, as outlined in Section \ref{abs}, has become a standard technique for the generation of tunable, monochromatic radiation up to 2~THz used for spectroscopic studies \citep{ARM2008,JAW2014,ZMM2015}. An extension of this technique to even higher frequencies ($2.5 -2.7$~THz) was demonstrated by \cite{PDM2011} at JPL, where a multistage multiplier chain originally developed as local oscillator source was used to measure the ground state rotational transition of HD \citep{DYP2011}, a prime target for astronomical observations with SOFIA/GREAT. Other promising THz radiation sources are quantum cascade lasers \citep[see][for recent reviews]{Wil2007,HEP2013}, and frequency down-conversion via photo-mixing \citep{BMN1995,HCB2008,DRS2015}, used, e.g., for the spectroscopy of OH and SH radicals up to 2.5~THz \citep{EMG2011,MEP2012}. The latter study combines high-resolution data with synchrotron-based Fourier-transform far infrared spectroscopy. This method, although not strictly high-resolution (limited to $\sim 20-30$~MHz linewidths), offers the advantage of acquiring very broadband spectra ($50 - 800$~cm$^{-1}$, i.e. $1.5 - 24$~THz) with high sensitivity. This makes it ideally suited to record spectra of low-lying vibrational modes of PAHs, even resolving the underlying rotational structure as demonstrated in the case of naphtalene, azulene, isoquinoline, and indole \citep{AAL2011,PGH2013,GP2014}.

\section{Rotational spectroscopy of molecular ions in trap instruments}

Many interstellar molecules are highly reactive radical and ionic species, rendering them importantant players in astrochemistry, but also very elusive for gas-phase spectroscopoic investigations. A widely-used method for the in-situ production of transient molecules are direct current (DC) or radio-frequency (RF) glow discharges.  Rotational spectra of the reaction products are obtained by absorption spectroscopy through the plasma, or by seeding them into a supersonic jet probed by CP-FTMW or cavity enhanced molecular beam (MB-) FTMW \citep[developed by][]{BF1981}. Although a very powerful method, with an ever increasing number of molecular radicals, cations and even anions studied this way, it also has several drawbacks. It is often difficult to produce high enough number densities of the target species for conventional spectroscopic methods, whereas on the other hand a large variety of other species are generated simultaneously, leading to very congested spectra aggravated by relatively high temperatures in the discharge. \\

\begin{figure}
   \centering
   \includegraphics[width=\hsize]{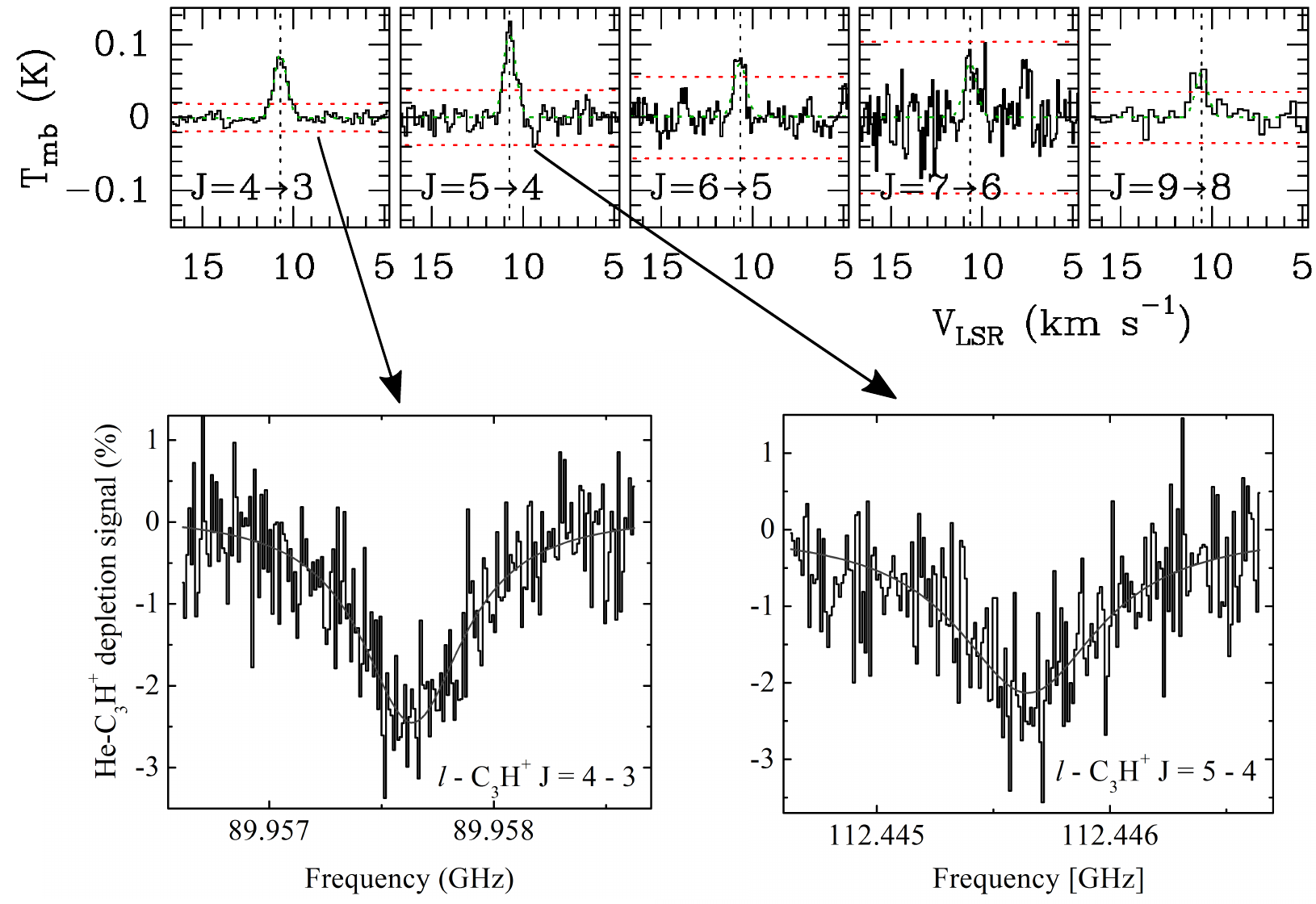}
      \caption{Laboratory identification of linear C$_3$H$^+$: Series of harmonic rotational emission lines towards the Horsehead PDR observed by \cite[][reproduced with permission {\textcopyright} ESO]{PGG2012} (top). Spectra of mass-selected linear C$_3$H$^+$ recorded with the method of state-selective helium attachment in a cryogenic ion trap, matching the astronomically observed lines (bottom).}
         \label{C3H+}
\end{figure}

In the Cologne laboratory astrophysics group we have in recent years followed a different approach and developed sensitive spectroscopic schemes based on mass-spectroscopic and cryogenic ion trapping techniques that can overcome these limitations. Molecular ions, generated by electron bombardment, are mass-selected prior to being stored and cooled in a cryogenic ($> 4$~K) multipole ion trap \citep{Ger1995,ABM2010,ABK2013}. A spectrum of the cold parent molecular ion is recorded by detecting a change in the ion composition as a function of the excitation wavelength. One of these very sensitive action spectroscopic schemes, requiring only a few hundred mass-selected stored ions, uses the promotion of a bimolecular endothermic reaction upon absorption of a resonant photon \citep[LIR - Laser Induced Reactions, ][]{SLR2002}. Working best for ro-vibrational transitions due to the higher photon energy absorbed, still rotational transitions can be predicted from these measurements to microwave accuracy by using a narrow-bandwidth OPO (optical parametrical oscillator) in the mid-infrared, as we have demonstrated in the case of CH$_2$D$^+$ \citep{GKK2013}. In certain cases rotational transition can be probed directly by LIR, e.g. the ground state transition of para-H$_2$D$^+$ \citep{ARM2008}, providing the spectroscopic basis for its recent detection and use as chemical clock in a protostellar core \citep{BSC2014}. Two-photon (IR-THz) experiments are also possible \citep[e.g. as shown for OH$^-$,][]{JAW2014} to highest accuracy. Another recently developed method uses the state selective attachment of helium to the cold stored molecular ions as action spectroscopic probe. This apparently very general method, working for rotational, ro-vibrational and even electronic spectroscopy, allowed us to measure the rotational spectrum of the linear C$_3$H$^+$ ion \citep{BKS2014}. Owing to the mass- and charge-selection inherent to the method, this study helped to settle a debate on the charge-state of the carrier of several astronomical lines observed towards the Horsehead PDR \citep{PGG2012,HFL2013,FHC2013}, unambiguously attributing them to the cationic C$_3$H$^+$ (see Fig. \ref{C3H+}).

\section{Conclusions}

Molecular spectroscopy is one of the driving fields in astrophysics as it allows to find new molecules in space
or known molecules in new environments like other galaxies, thus literally opening new rooms.
Often these molecules are carrying information on specific astronomical environments where they reside.
They are not just observers of what happens in these environment. In many cases 
like the examples discussed here these molecules are active players
in the underlying astrochemistry. 

As has been summarized in this short review, laboratory scientists currently focus
on specific classes of molecules to answer pestering questions of astrophysics: \\
i) How complex do molecules grow in space?
Is it possible to find, e.g., aminoacids? \\
ii) How do such molecules form under the extreme conditions in space?
What are the roles of gas phase or heterogeneous processes on grains and ices?\\
iii) What are the abundance ratios of isotopologues in space and how could this 
information be used to infer the chemical history of the various environments in space?\\
iv) Which abundant or astrochemically important molecules are missing in the list of interstellar molecules?

Recording the laboratory spectra of these and other molecules is associated with tremendous technical challenges.
The most pressing are creating sufficient amounts of the molecule of interest
and isolating them from other species contaminating the spectra.
Our ability to address these questions with the help of new techniques like spectroscopy in traps determines the
speed at which interesting new molecules like cations, anions, radicals or 
other more exotic molecules can be found in observations.
Recording spectra at different temperatures or predicting them at arbitrary temperatures remains difficult.
Understanding the underlying energy structure of the many thousand transitions associated with COMs
and other complex molecules is an important task. Predicting spectra of molecules with many electronic states like
metal containing species to spectroscopic precision is a challenge to quantum chemistry.
As a consequence, progress in molecular spectroscopy will be made through innovative experimental methods as well as
computational efforts which deal with the different degrees of complexity of the spectra which lie ahead.
Automated analysis and interpretation of the wealth of data from the laboratory and space alike 
will be the most demanding task for the coming years.



\end{document}